\documentclass[11pt,twoside]{jgsp}
\usepackage{amsfonts}
\usepackage{amsmath}

\def\aut#1#2{
{\small
\noindent
\parbox[t]{6.5cm}{#1}
 \hfill
\parbox[t]{6.5cm}{#2}
}
}

\begin{document}
\thispagestyle{plain}
\title{Some solutions with torsion in Chern-Simons gravity and observable
effects}
\author{Fabrizio Canfora}
\date{}
\maketitle
\begin{abstract}
It is shown that in five dimensional Einstein-Gauss-Bonnet
theories exact vacuum solutions with non-vanishing torsion can be
constructed.

Some possible observational effects related to neutrino
oscillations are pointed out.

In the theory of continuum media (in which suitable defects can be
described by localized non vanishing torsion) "the gravitational
intuition" is a rather useful tool to describe the physical
effects of such defects. A possible astrophysical application is
shortly described.
\end{abstract}

\label{first}

\section{Introduction}

Finding the correct quantum theory of gravity is a major problem in
theoretical physics. Although there are very strong proposals such as
superstring theory and loop quantum gravity, the final answer has not been
found yet. Thus, it is worth to explore new possibilities. Among the many
proposals available in the literature, the possibility to have gravitational
actions different from the Einstein-Hilbert one and with non vanishing
torsion is worth to be explored. The most natural generalizations of the
Einstein-Hilbert action are the so called Lovelock action \cite{Lovelock}:
Lovelock actions are the most general covariant actions leading to second
order equation for the metric (it is a very difficult task to provide one
with a complete list of references on the subject; a nice review on the
black hole solutions is \cite{Gaston}). Unlike the four dimensional case in
which torsion vanishes, in higher dimensional Lovelock theories torsion may
represent propagating degrees of freedom \cite{TZ99} \cite{Za05}. However,
up to now the possible dynamical role of torsion has not been taken into
account too seriously.

In this paper an intriguing analogy between BPS states in QFT and gravity
with torsion will be explored. This analogy allows to find interesting
solutions in five dimensional Einstein-Gauss-Bonnet gravity in which the
torsion looks like topological excitations and may improve the stability of
torsion-free solutions. Furthermore, torsion could led to observable effects
both in high energy physics and in solid state physics.

The paper is organized as follows: in the second section a very short review
of BPS states in field theory is presented. In the third section, a general
introduction to gravitational models with torsion is described. In the
fourth section the analogy of BPS states in field theory and gravitational
model with torsion is presented and how to construct solutions of
Chern-Simons gravitational theories with torsion starting from solutions
without torsion is discussed. In the fifth section, some observable effects
in high energy physics and in solid state physics are discussed. Eventually,
some conclusions are drawn.

\section{Very Short Review on\ BPS in Field Theory}

In this section it will be shortly resumed BPS states in field theory with
attention on vortices. Because of the enormous number of papers on this
subject, it is a hopeless task to provide with a complete list of
references, the main reference on which is based the present paper is the
detailed review \cite{OW99} \cite{To05}.

A suitable gauge theoretical action able to give rise to vortex-like
solutions is%
\begin{equation*}
S=\int d^{4}x\left[ \frac{1}{4}F_{\mu \nu }F^{\mu \nu }+\left\vert D_{\mu
}\phi \right\vert ^{2}-\frac{\lambda }{4}\left( \left\vert \phi \right\vert
^{2}-v^{2}\right) ^{2}\right]
\end{equation*}%
where $F_{\mu \nu }$ is the field strength of an Abelian $U(1)$
connection, $ \phi $ is a complex $U(1)$-charged scalar field
(whose charge is $e$) with the standard symmetry breaking
potential whose minima are of the form
\begin{equation}
\phi _{vac}=v\exp (i\alpha )  \label{vac1}
\end{equation}%
where $\alpha $ is a real phase. The energy of a static vortex
configuration (in which the vortex is along the third axis) is
\begin{equation*}
E_{AH}=\int d^{2}x\left[ \frac{1}{2}B^{2}+\left\vert \overrightarrow{D}\phi
\right\vert ^{2}+\frac{e^{2}}{2}\left( \left\vert \phi \right\vert
^{2}-v^{2}\right) ^{2}\right] \geq v^{2}\left\vert \Phi _{B}\right\vert
\end{equation*}%
$\Phi _{B}$ being the magnetic flux. For any finite energy solution $\phi $
approaches the vacuum manifold at infinity. Therefore the finite energy
solutions are classified by their winding number $N$ (also called vorticity)
which counts the number of times the phase of $\phi $ winds around the
spatial circle at infinity%
\begin{equation*}
\phi \underset{r\rightarrow \infty }{\rightarrow }v\exp (iN\theta ).
\end{equation*}%
The above winding number labels homotopy classes. The vorticity is related
to the magnetic flux $\Phi _{B}$ since%
\begin{equation*}
\left\{ \overrightarrow{D}\phi \underset{r\rightarrow \infty
}{\rightarrow } 0\right\} \Rightarrow eA_{i}\sim -i\partial
_{i}\log \phi =N\partial _{i}\theta \Rightarrow \Phi _{B}=2\pi N.
\end{equation*}%
The Bogomol'nyi equations can be deduced assuming that the bound is
saturated:%
\begin{equation*}
B\mp e\left( \left\vert \phi \right\vert ^{2}-v^{2}\right) =0,\quad \left(
D_{1}\pm iD_{2}\right) \phi =0.
\end{equation*}
Such equations (which are linear relations among the magnetic part
of the curvature and the square of the Higgs field) will be very
important to formulate the analogy and to establish a dictionary
between "BPS gauge theoretical objects" and "gravitational
torsional objects". It is worth to stress that also in the case of
monopoles in non Abelian Yang-Mills-Higgs theory the Bogomol'nyi
equations are linear relations among the magnetic components of
the curvature and the covariant derivative of the Higgs field of
the type
\begin{equation*}
B^{a}=\pm \left( D\phi \right) ^{a}
\end{equation*}
where $B^{a}$ is the non abelian magnetic field. An important point is that
often (in this case as well as in the case of monopoles) Bogomol'nyi
equations can be formulated as self-duality conditions of a "bigger
curvature" of a "bigger connection" in which the gauge connection and the
Higgs field fit together \cite{OW99} \cite{To05} in such a way that the
self-duality conditions for the bigger curvature are the precisely the above
described linear relations among the gauge field strength, the Higgs field
and its covariant derivative.

\section{Short Review in Gravity with Torsion}

Here a very short introduction to gravitational actions with torsion will be
provided (see the detailed reviews \cite{TZ99} \cite{Za05}).

In the first order formalism the Einstein-Hilbert action in four
dimensions can be written as follows:
\begin{equation*}
S_{EH}=\int\varepsilon_{abcd}\left( \alpha R^{ab}e^{c}e^{d}+\beta
e^{a}e^{b}e^{c}e^{d}\right)
\end{equation*}
where
\begin{align*}
e^{a} & =e_{\mu}^{a}dx^{\mu},\quad R_{b}^{a}=\left( d\omega+\omega
\wedge\omega\right) _{b}^{a}, \\
\omega_{b}^{a} & =\omega_{b\mu}^{a}dx^{\mu},\quad
g_{\mu\nu}=\eta_{ab}e_{\mu}^{a}e_{\nu}^{b}
\end{align*}
$e^{a}$ is the vielbein, $\omega_{b}^{a}$ the spin connection,
$g_{\mu\nu}$ is the spacetime metric, $\eta_{ab}$ the Minkowski
metric in the vielbein indices and $\alpha$ and $\beta$ are
related in a obvious way with the Newton and the cosmological
constants. It is manifest in this way that the vielbein indices
$a,b,..$ behave as internal gauge indices. Since the Kaluza-Klein
idea and with the advent of string theories the possibilities to
have extradimensions comes into play. Lovelock \cite{Lovelock}
showed that it is indeed possible to have gravitational Lagrangian
in higher dimensions giving rise to second order equations of
motions for the metric, the Lagrangian being of the form
\begin{align*}
I_{D} & =\kappa\int\sum_{p=0}^{\left[ D/2\right] }\alpha_{p}L^{(D,p)}, \\
L^{(D,p)} &
=\varepsilon_{a_{1}..a_{D}}R^{a_{1}a_{2}}..R^{a_{2p-1}a_{2p}}e^{a_{2p+1}}..e^{a_{D}}.
\end{align*}
In $D=5$ (on which we will focus from now on) the so called
Einstein-Gauss-Bonnet action \cite{Lovelock} reads%
\begin{equation}
I=\kappa\int d^{5}x\sqrt{g}\left( R-2\Lambda+\alpha\left( R^{2}-4R_{\mu\nu
}R^{\mu\nu}+R_{\alpha\beta\gamma\delta}R^{\alpha\beta\gamma\delta}\right)
\right) \ ,  \label{Itensor}
\end{equation}
where $\kappa$ is related to the Newton constant, $\Lambda$ to the
cosmological term, and $\alpha$ is the Gauss-Bonnet
coupling\footnote{ The relationship between the constants
appearing in Eqs (\ref{Itensor}) and ( \ref{action}) is given by
$\alpha=\frac{c_{2}}{2c_{1}}$, $\Lambda =-6\frac{ c_{0}}{c_{1}}$,
$\kappa=2c_{1}$ .}. The action (\ref{Itensor}) in terms of
differential forms reads

\begin{equation}
I=\int\epsilon_{abcde}\left(
\frac{c_{0}}{5}e^{a}e^{b}e^{c}e^{d}e^{e}+\frac{
c_{1}}{3}R^{ab}e^{c}e^{d}e^{e}+c_{2}R^{ab}R^{cd}e^{e}\right) \ .
\label{action}
\end{equation}

The field equations read
\begin{equation}
\mathcal{E}_{e}:=\left(
c_{o}e^{a}e^{b}e^{c}e^{d}+c_{1}R^{ab}e^{c}e^{d}+c_{2}R^{ab}R^{cd}\right)
\epsilon_{abcde}=0\ ,  \label{equationcurvature}
\end{equation}
and
\begin{equation}
\mathcal{E}_{ab}:=T^{c}\left( c_{1}e^{d}e^{e}+2c_{2}R^{de}\right)
\epsilon_{abcde}=0\ ,  \label{equationtorsion}
\end{equation}
while the torsion is
\begin{equation*}
T^{a}=De^{a}.
\end{equation*}

In the vanishing torsion sector, the field equations
(\ref{equationtorsion}) are trivially fulfilled, and Eq.
(\ref{equationcurvature})\ reduces to the standard one in the
second order formalism. An interesting characteristic of the
Einstein-Gauss-Bonnet theory is that, unlike in the
Einstein-Hilbert case, torsion is generically different from zero
\cite{TZ99} (even if many authors put torsion equal to zero by
hand). Generically, the field equations (\ref{equationtorsion})
impose very strong constraints on the torsion (which can be found
by taking the covariant derivative of Eq.
(\ref{equationcurvature}) and comparing the result with Eq.
(\ref{equationtorsion}
)):%
\begin{equation}
e^{a}R^{bc}T^{d}\epsilon _{abcde}=0,\ \ \ e^{a}e^{b}e^{c}T^{d}\epsilon
_{abcde}  \label{constrtors}
\end{equation}
so that, to the best of author's knowledge, no exact solution with
torsion has been found in the generic Lovelock case up to recent
times. As it will be described in the next section, a nice analogy
with BPS states in field theory suggests a natural ansatz for the
torsion which solves automatically the constraints in Eq.
(\ref{constrtors}) allowing to construct exact solutions with
torsion \textit{in the non-Chern-Simons case} \cite{CGW07}.
However, there exist special values of the Gauss-Bonnet coupling
which enhance the gauge symmetry of the generic Lovelock
actions\footnote{ Indeed, for such special values, the Lagrangian
can be written as a Chern-Simons form \cite{Chamseddine}. In the
Chern-Simons case, some exact solutions with torsion have been
constructed in \cite{Ar06}: however, it seems that in these
solutions torsion is pure gauge.} \cite{TZ99} "killing" the
constraints in Eq. (\ref{constrtors})
\begin{equation}
c_{2}=\frac{c_{1}^{2}}{4c_{0}}\ ,  \label{choice}
\end{equation}%
leading to the uniqueness of the maximally symmetric vacuum \cite{BH-Scan}.
In this case, a supergravity theory is available and the "BPS form" of the
torsion provides one with black hole solutions with a half-BPS ground state
\cite{CGTW07}.

\section{Gauge-Theoretical Interpretation of Torsion}

A useful trick which helps in dealing with torsion in four
dimensions is available in the presence of a non vanishing
cosmological constant: one can define the following "higher order"
connection
\begin{equation}
W^{AB}=\left[
\begin{array}{cc}
\omega^{ab} & -\frac{e^{a}}{l} \\
-\frac{e^{b}}{l} & 0
\end{array}
\right]  \label{conn1}
\end{equation}
where $a,b=1,..,4$ and $A,B=1,..,5$ in such a way that the "higher
order" curvature is
\begin{equation}
F^{AB}=\left[
\begin{array}{cc}
R^{ab}-\frac{e^{a}\wedge e^{b}}{l^{2}} & \frac{T^{a}}{l} \\
-\frac{T^{b}}{l} & 0
\end{array}
\right]  \label{conn1.5}
\end{equation}
and the "higher dimensional" topological term reads%
\begin{equation}
F^{AB}\wedge F_{AB}=R^{ab}\wedge R_{ab}+\frac{2}{l^{2}}N.  \label{conntop}
\end{equation}
It is worth to stress here that this is also a standard trick in the theory
of BPS states which explains why the BPS equations are much simpler than the
full Yang-Mills equations: the point is that often one can incorporate the
Higgs field and the Yang-Mills connection into a unique higher connection in
such a way that the BPS equations are self-duality conditions for the higher
curvature (see, for example, \cite{OW99}). An immediate consequence is that%
\begin{equation}
\int\frac{2}{l^{2}}N=8\pi^{2}\times n,\quad n\in \mathbb{Z} ,
\label{quacotor}
\end{equation}
that is, two times the integral of the Nieh-Yan density \textit{divided by
the square of the cosmological length} is $8\pi^{2}$ times an integer (see,
for instance, \cite{CZ97}).

The previous trick to incorporate torsion, vielbein and curvature in a
"bigger curvature" is standard in studying BPS states in QFT. This analogy
suggests that, in gravity, the vielbein and torsion could play the role of
the Higgs field and its covariant derivatives in QFT. In other words, one
would like to give a precise meaning to the following table:

\begin{equation*}
\begin{array}{cc}
Gravity & Gauge\quad Theory \\
e^{a} & \phi^{a} \\
T^{b} & \left( D\phi\right) ^{b} \\
\alpha R^{ab}+\frac{\beta}{2}e^{a}e^{b} & F^{ab}%
\end{array}
\end{equation*}
where $\alpha$ and $\beta$ are suitable constants. In QFT the BPS
equations involve typically linear relations (such as higher
dimensional self-duality conditions) among $D^{a}\phi$ and
$F^{ab}$ in which $\phi$ can enter quadratically (as, for
instance, in the vortex case). Thus, a natural ansatz is
\begin{equation}
T^{c}=f^{cab}\left( \alpha R^{ab}+\beta e^{a}e^{b}\right)  \label{gravBPS1}
\end{equation}
where $f^{cab}$ is a three index tensor in the "gravitational
gauge group", $ \alpha$ and $\beta$ are two constants.

However, a genuine invariant tensor (that is, $f^{abc}$ in Eq.
(\ref{gravBPS1})) in the Lorentz group does not exist in general.
As it is well known (see, for instance, \cite{We96}) some breaking
of the vacuum symmetries is needed. In this case, a tensor
$f^{abc}$ which makes consistent Eq. (\ref{gravBPS1}) can be
constructed using the broken generator(s). The simplest invariant
tensor with three indices is the Levi-Civita symbol $\varepsilon
^{abc}$. In the five-dimensional case (which has been analyzed in
\cite{CGW07}\ and \cite{CGTW07}) this corresponds to consider
solutions which are a direct or a warped product of a two
dimensional manifold $M_{2}$\ and a three dimensional manifold
$M_{3}$ of constant curvature on which the torsion lives:
\begin{equation}
T^{c}=K\left( x^{1},x^{2}\right) \varepsilon ^{cab}e^{a}e^{b}
\label{gravBPS1.1}
\end{equation}
where $K$ may depend on the coordinates ($x^{1}$, $x^{2}$) of
$M_{2}$ and the vielbeins $e^{a}$ and the indices $a$, $b$ and $c$
refer to $M_{3}$. Remarkably enough, this ansatz automatically
solves both constraints in Eq. ( \ref{constrtors}).

\subsection{The non-Chern-Simons case}

For the first time, the above ansatz (\ref{gravBPS1.1}) allowed to find
\textit{exact vacuum solutions with torsion} in five dimensional
Einstein-Gauss-Bonnet gravity for a suitable choice of the constants
appearing in the action which does not correspond to a Chern-Simons theory
\cite{CGW07}. The metric is given by
\begin{equation*}
ds^{2}=ds_{AdS_{2}}^{2}+ds_{S^{3}}^{2}.\,
\end{equation*}
The torsion is introduced onto the $M_{3}$ in such a way as to respect the
symmetry:
\begin{equation*}
T^{a}=\tau\,\epsilon_{abc}e^{b}e^{c}\,.
\end{equation*}

By imposing the relation
\begin{equation}
c_{1}^{2}=12c_{0}c_{2}  \label{nocs}
\end{equation}
(which, interestingly enough, does not correspond to the
Chern-Simons case) one gets an exact vacuum solution with torsion
in Einstein-Gauss-Bonnet gravity whose curvatures are
\begin{gather}
T_{c}=\pm\sqrt{\frac{1}{D^{2}}+\frac{1}{3l^{2}}}\
\epsilon_{abc}\,e^{b}e^{c}\,,  \label{Answer_Torsion} \\
R^{01}=-\frac{1}{l^{2}}e^{0}e^{1}\,,\qquad
R^{ab}=-\frac{1}{3l^{2}} \,e^{a}e^{b}\,,
\end{gather}
provided that $c_{2}/c_{1}$ and $c_{0}/c_{1}$ are positive. It is worth to
note that the torsion is bounded from below for finite sized sphere and AdS
length scale. This means that there is no continuous zero torsion limit.

Furthermore, if one defines the Hodge dual of the three form $T$
\begin{align*}
T & \equiv T_{abc}e^{a}\wedge e^{b}\wedge e^{c}=3!\tau\,e^{2}\wedge
e^{3}\wedge e^{4}, \\
\ast T & =-2!\tau e^{0}\wedge e^{1},
\end{align*}
then
\begin{equation*}
DT=0,\qquad D\ast T=0
\end{equation*}
so that, by defining $F\equiv\ast T$, $F$ is seen to obey the source-free
Maxwell equations, making manifest the close resemblance with the
electromagnetic field of the Bertotti-Robinson solution \cite{BR}.

\subsection{The Chern-Simons case}

Let us now shortly describe the Chern-Simons case in which%
\begin{equation*}
c_{1}^{2}=4c_{0}c_{2}.
\end{equation*}
In the case of the five-dimensional Chern-Simons gravity, spherically
symmetric black hole without torsion are well known:
\begin{equation*}
ds^{2}=-\left( \frac{r^{2}}{l^{2}}-\mu\right)
dt^{2}+\frac{dr^{2}}{\frac{r^{2}}{l^{2}}-\mu}+r^{2}d\Sigma_{3}^{2}\ ,
\end{equation*}
and are quite similar to three-dimensional BTZ black-hole \cite{Banados:1992wn}, \cite{Banados:1992gq}. In both cases, when the
mass parameter $\mu$ is $\mu=-1$ the solution is $AdS_{3}$ and
$AdS_{5}$. When the mass parameter is greater than zero one has a
black hole and when $ -1<\mu<0$, timelike naked singularities
appear. When torsion vanishes and without matter field, in the
five dimensional case the zero mass black hole breaks all the
supersymmetries \cite{SUSY}. While in three dimensions the $
\mu=0$ is the half-BPS "black hole vacuum". If torsion is
introduced in $ \Sigma_{3}$ as follows
\begin{equation}
T^{a}=-\frac{\delta}{r}\epsilon^{abc}e_{b}e_{c}\ ,  \label{Torsion-BH}
\end{equation}
the above metric still represents a five-dimensional black hole
\cite{CGTW07}. Furthermore, when the constant $\delta$ is equal
to one (or minus one), it can be shown that the solution with
$\mu=0$ is half-BPS exactly in the same way as it happens in three
dimensions (\textit{something impossible without torsion}). Thus,
torsion improves the stability analysis of the theory.

\subsection{A physical Interpretation}

The nice features of the results in \cite{CGW07} and \cite{CGTW07} (which
have been resumed in the previous subsections) is that in five (and,
presumably, higher dimensions) one can construct with a very natural recipe
exact \textit{vacuum} solutions with non-trivial geometrical structures
which, in four dimensions, could only be obtained coupling gravity to
suitable matter fields. The reason is that, being the total connection the
sum of the Riemannian part plus the contorsion, one can manage to choose the
contorsion in such a way to exactly cancel (a suitable part of) the purely
Riemannian connection (as it has been done in \cite{CGTW07}). This is quite
similar to what happens, for instance, in the case of the extreme charged
black-hole in Einstein gravity: the electromagnetic connection cancels part
of the Levi-Civita connection leading to an half-BPS solution. This means
that, in higher dimensions, the gravitational interaction is much more
complex allowing, in some cases, to "emulate" matter fields much more
effectively than in Einstein gravity: in other words, in higher dimensions
gravity can be more "repulsive" than in four dimensions providing one with
the possibility to shed new light on some of the open question in cosmology.
One can see this as follows: from a particles physics perspective, the
Riemannian part of the gravitational interactions can be represented by a
spin-two particle, the graviton, which generates basically an attractive
interaction. In four dimensions, this is the only possibility compatible
with the requirements of general covariance and with second order equations
of motion for the metric. In higher dimensions torsion is not zero anymore
so that also a non-Riemannian part of the gravitational field appears. In
many cases, the torsion generates a repulsive interaction: for instance, in
the previous subsections it has been shown that torsion may behave as an
electromagnetic field generating in vacuum a Bertotti-Robinson-like
solution. Furthermore, torsion may also stabilize the spherically symmetric
Chern-Simons black-hole preventing a decay into naked singularities. For
this reason, the role of torsion is worth to be further investigated.

\section{Observable effects}

\bigskip In this section some observable effects in high energy physics and
solid state physics related to the above discussion will be analyzed.

\subsection{High energy physics}

If the Einstein-Hilbert Lagrangian is the reduction of some higher
dimensional model (indeed, Chern-Simons theories are very natural candidates
\cite{AVZ06}) the torsion generically does not vanish and propagates \cite%
{BGH96}. One should consider the possibility that somewhere in the universe
torsion could be present. Let us consider the case of Chern-Simons theory
(related to the (A)dS vacuum) in which torsion is non zero only along two
direction (say, $1$ and $2$). One can consider four dimensional cosmological
models of Einstein equations as coming from a suitable Chern-Simons
solutions.

An interesting effect of non vanishing torsion is related to neutrinos
oscillations. In \cite{AD01} it has been shown that the torsion can
contribute to neutrino oscillation along a path of length $\Delta r$ (among
neutrinos of different polarization):%
\begin{equation}
\left( \Delta\Phi\right)
_{T}=\Phi^{+}-\Phi^{-}\approx\frac{1}{l^{\prime}} \Delta r
\label{torneu}
\end{equation}
where the $+$ and the $-$ refer to the up and down polarizations,
$l$ is the coefficient in front of the torsion term in the
covariant derivative. The contribution of torsion to oscillations
can be understood because for Fermion, in rough terms, the
coupling with torsion is similar to the coupling with mass. The
authors of \cite{AD01} took as $l^{\prime}$ the Planck length
(arguing that the torsion, being a quantum effects, should appear
at the Planck scale): the result was that the contribution of the
torsion is 55 order of magnitude less than the field theoretical
one $\left( \Delta\Phi\right) _{QFT}$
\begin{align*}
\left( \Delta\Phi\right) _{QFT} & \approx\frac{c^{4}\Delta
m^{2}}{2E\hbar c}
\Delta r, \\
\frac{c^{4}\Delta m^{2}}{2E\hbar c} & \approx10^{-9}\left[ m\right] ^{-1}.
\end{align*}
It is much more natural from the theoretical point of view (see, for
instance, \cite{TZ99}\ \cite{Za05}) that the constant $l^{\prime}$ is
related to the cosmological constant. In such a case, the contribution of
the torsion is six order of magnitude larger then the observed one. Thus, in
this case also it seems not possible to explain neutrino oscillations with
torsion.

It cannot be excluded that in a more refined setting one could get an order
of magnitude closer to the observed one. The results of \cite{AD01} follow
from two assumptions: the first is that "the torsion coupling constant" is
the Planck length\footnote{%
It has been already discussed that it is almost mandatory from the geometric
point of view to assume that "the torsion coupling constant" is the
cosmological length.}. The second is that the torsion is related to the
Dirac fields describing the neutrinos as dictated by the Einstein equations
and, moreover, it is assumed that the torsion has not a dramatic dependence
on the spacetime coordinates\footnote{%
In practice, in \cite{AD01} the possible spacetime dependence of the torsion
is neglected.} so that, in order to estimate the effects of the torsion it
is enough to take into account the order of magnitude of the torsion
coupling constant. It has been shown that there exist solutions in five
dimensional Chern-Simons theory in which one has non trivial torsion along
some directions and that a non trivial spacetime dependence is possible. In
the present formalism (assuming for simplicity that there is only one
deformed tetrad field), the phase factor arising in Eq. (\ref{torneu}) of
\cite{AD01} is related to the integral
\begin{equation*}
\frac{1}{l}\int \limits_{\gamma}\delta e_{\mu}^{1}dx^{\mu}
\end{equation*}
where $\delta e_{\mu}^{1}$ is the deformation of tetrad field responsible of
the non vanishing of the torsion and $\gamma$ is a spacelike geodesic
joining the two points among which one wants to compute the oscillations.
One obtains the results of \cite{AD01} in the case in which there is no
dependence on the coordinates. However, at this stage of the analysis it is
not clear how to determine the torsion from first principles since in
Chern-Simons theory there is a huge freedom allowing for too many different
deformations of the tetrad.

\subsection{Solid state physics}

In the theory of elasticity the analysis of defects can be conveniently
formulated (see \cite{KV92} \cite{FS93} \cite{Ka04} \cite{CV06}) by
describing the medium as a continuum geometry with localized torsion. In
particular, torsion singularities located on parallel lines appears to be
well suited to describe dislocations \cite{KV92} \cite{FS93} while the
geometry of cosmic strings can describe rather well \cite{CV06}
disclinations present in various type of materials.

The Dirac equation in presence of torsion can be written as
follows
\begin{equation}
i\Gamma^{\mu}\left(
\partial_{\mu}-\frac{\omega_{\mu}^{ab}}{4}\Gamma _{ab}+
\frac{1}{2l}e_{\mu}^{a}\left( (i)^{n}\Gamma_{a}\right) \right)
\Psi(x)=0 \label{dirtor}
\end{equation}
where $\omega_{\mu}^{ab}$ is the Levi-Civita connection,
$\Gamma^{a}$\ are the flat Dirac matrices, $\Gamma^{\mu}$\ the
curved ones\footnote{The above Dirac equation is an effective
equation because it can only describe the low energy electronic
excitations near the Fermi surface. Once the propagator is known,
it is possible to compute the electronic density around defects.},
$\Psi$\ is the Dirac spinor, $l$ is a length scale and $n$ can be
zero or one (as it will be discussed in a moment together with the
physical meaning of length scale $l$). The added term is quite
natural in a curved spacetime and should not be neglected. When
dealing with standard gravitational models, it can be gauged away:
\begin{align*}
\partial_{\mu}-\frac{\omega_{\mu}^{ab}}{4}\Gamma_{ab}+\frac{1}{2l}%
e_{\mu}^{a}\left( (i)^{n}\Gamma_{a}\right) &
\rightarrow\partial_{\mu}-\frac{\omega_{\mu}^{ab}}{4}\Gamma_{ab} \\
\Psi(x) & \rightarrow P\exp\left[ i\int^{x}e_{\mu}^{a}(y)\left(
(i)^{n}\Gamma_{a}\right) dy^{\mu}\right] \Psi(x)
\end{align*}
where $P$ is the path ordering. The integral in the phase factor
does not depend on the path because the covariant derivative of
the tetrad $ e_{\mu}^{a}$ vanish and therefore it has no physical
effects. In the presence of dislocations this is not true anymore
since the \textit{torsion does not vanish}. When the torsion is
localized along the $z$ axis the Aharonov-Bohm phase factor $\Phi$
is
\begin{equation}
\Phi=\exp\left[ \frac{i}{2l}\int\int dxdyT\right] .  \label{fafa}
\end{equation}
The integral in the exponent has a very nice physical
interpretation: it is the so called \textit{Burgers vector}
$\overrightarrow{b}$\ which is proportional to the the integral of
the flux if the torsion through a surface \cite{KV92}:
\begin{equation*}
\int\int_{S}dx^{m}\wedge dx^{n}T_{mn}^{i}=-b^{i}.
\end{equation*}
Therefore, the length scale $l$ in a solid state physics context has to be
interpreted as the mean "size" of the Burgers vector%
\begin{equation*}
2l\sim\left\vert \overrightarrow{b}\right\vert .
\end{equation*}
Such a vector is very important to describe the physical effects
of edge dislocations which are rather common in
nature\footnote{"Pictorially" such defects arise when one "cuts"
some pieces of material and then glue the boundary in a suitable
way with or without adding extra pieces of material obtaining a
much higher or a much lower density. Non vanishing torsion can
describe both wedge dislocations (which are rather uncommon) and
edge dislocations (which are often encountered in practical
applications) \cite{Ka04}.}. Another interesting possibility has
been proposed in \cite {Ba98} in which topological defects
manifest themselves through anomalous conductivities.

The quantum effects of dislocations can manifest themselves also
by computing the propagator in an effective three dimensional
gravitational field with torsion localized along some spacelike
line. One can think to a three dimensional piece of material in
which the interesting phenomena occur in a two dimensional planes
orthogonal to the dislocation. Let us consider the effective Dirac
equation Eq. (\ref{dirtor}) describing the low energy electronic
excitations in the field of a "vortex of torsion" along the $z$
axis (but without other defects near the dislocation):
\begin{equation}
i\Gamma^{\mu}\left( v_{F}\partial_{\mu}+\frac{(i)^{n}}{2l}e_{\mu}^{a}\left(
\Gamma_{a}\right) \right) \Psi(x)=0  \label{densel}
\end{equation}
where $v_{F}$\ is the Fermi velocity of the material (which,
henceforth will be set equal to one). To proceed, one has to
assume that the "torsional" term can be treated as a small
perturbation\footnote{In the case of the defects analyzed in
\cite{CV06} such approximation gives results in very good
agreement with observations.}.

Besides the obvious term, the tetrad fields and the curved metric
also manifest themselves through the curved Dirac matrices. In the
present case, one is analyzing an effective two-dimensional
situation in which only the third tetrad field is deformed in such
a way to achieve a "vortex of torsion" along the $z$-axis and the
meaningful dynamics is in the $x-y$ plane so that the explicit
dependence on $z$ of the Dirac wave function can be neglected. The
structure of the spatial tetrad fields is
\begin{equation*}
e^{1}=dx,\quad e^{2}=dy,\quad e^{3}=dz+a_{i}dx^{i},\quad i=1,2=x,y
\end{equation*}
(where the explicit form of $a_{1}$ and $a_{2}$ is not important
in this moment) so that the inverse tetrads read
\begin{align*}
e_{a} & =e_{a}^{(0)}+\delta e_{a}^{(0)}:\quad e_{3}=\partial_{z} \\
e_{1} & =\partial_{x}-a_{1}\partial_{z},\quad
e_{2}=\partial_{y}-a_{2}\partial_{z}.
\end{align*}
The further terms $\delta\left( \Gamma^{\mu}\partial_{\mu}\right) $ in Eq. (
\ref{densel}) coming from the curved Dirac matrices are
\begin{equation*}
\delta\left( \Gamma^{\mu}\partial_{\mu}\right) =-\left( a_{1}\Gamma
^{1}+a_{2}\Gamma^{2}\right) \partial_{z}.
\end{equation*}
In the approximation in which the effective dynamics is two dimensional, the
derivative with respect to $z$ can be neglected. Therefore, the only further
term in Eq. (\ref{densel}) with respect to a flat Dirac equation is the one
in which it explicitly appears the tetrad field multiplied by $\frac{(i)^{n}%
}{2l}$.

Thus, when the effective dynamics is two-dimensional (so that the dependence
on $z$ can be neglected), the effect of dislocations can be described in
exactly the same way as the effects of disclinations. Thus, for instance, if
one manages to compute the total density of states $\rho (\omega )$ of the
electrons localized near a disclination then one can use the same formalism
to compute the density of electrons near a dislocation. In order to proceed
in the computation, one needs the Fourier transform of $\delta e_{\mu }^{3}$
(which, in the present geometric perspective, is the "source of torsion").\
Dislocations can be described as torsion localized along some axis (say, the
$z$ axis).\ In the case in which the geometry is locally flat besides the
dislocation itself, the suitable form of $\Gamma _{3}\delta e_{i}^{3}$ able
to describe torsion localized along the $z$ axis is formally the same as the
one needed to describe a thin magnetic flux along the same axis:%
\begin{equation*}
\frac{\Gamma _{3}\delta e_{j}^{3}}{2l}=A_{j}^{Vort}=\frac{\Theta }{2\pi }
\varepsilon _{3jk}\frac{x^{k}}{r^{2}}
\end{equation*}%
where $\Theta $ represents the strength of the torsion (which, as it has
been already stressed, is proportional to the modulus Burgers vector):%
\begin{equation*}
\Theta =\int\limits_{\gamma }A_{j}^{Vort}dx^{j}
\end{equation*}%
$\gamma $ being a closed curve around the origin in the $x-y$ plane. After
this identification the computations to obtain the electron density near the
defect may follow the line of \cite{CV06}\footnote{%
In which, however, there is a mistake. I thank M. A. H. Vozmediano\ for
pointing out this to me. However, the important point of the present
discussion is that in principle one can get the effects of dislocations in
two space dimensions from the analogous computation for a disclination.
Thus, the technical mistake in \cite{CV06} does not affect this possibility.}%
.

Because of its reach geometrical structure, the phenomenological role of
torsion in solid state physics is worth to be further investigated.

\section{Conclusion}

It has been shown that in higher dimensional gravitational Chern-Simons
theories there exist solutions with non vanishing torsion which rapidly
approaches to zero outside suitable spacetime regions. The possible
contribution of the torsion to neutrino oscillations has been already
discussed. However, this formalism clarifies that if one correctly takes
into account the geometric nature of torsion the contribution of torsion to
neutrino oscillations could be rather closer to the observed value. This
scheme has also interesting applications in solid state physics.
Dislocations (rather common defects in solid state physics) could be
observable both via Aharonov-Bohm type experiments.

\section*{Acknowledgements}

The author would like to thank J. Zanelli for many suggestions and
continuous encouraging, Andres Anabalon, Alex Giacomini, Ricardo Troncoso
and Steven Willison for many stimulating discussions and important
bibliographic suggestions. This work was supported by Fondecyt grant
3070055. The Centro de Estudios Cient\'{\i}ficos (CECS) is funded by the
Chilean Government through the Millennium Science Initiative and the Centers
of Excellence Base Financing Program of Conicyt. CECS is also supported by a
group of private companies which at present includes Antofagasta Minerals,
Arauco, Empresas CMPC, Indura, Naviera Ultragas and Telef\'{o}nica del Sur.
CIN is funded by Conicyt and the Gobierno Regional de Los R\'{\i}os.

\aut{Fabrizio Canfora\\
Area de Fisica Teorica \\
Centro de Estudios Cientificos (CECS)\\
Valdivia, Chile.\\
{\it E-mail address}:\\
 {\tt canfora@cecs.cl}}  {}

\label{last}

\end{document}